\documentclass[lettersize,journal]{IEEEtran}
\usepackage{amsmath,amsfonts}
\usepackage{algorithmic}
\usepackage{algorithm}
\usepackage{array}
\usepackage[caption=false,font=normalsize,labelfont=sf,textfont=sf]{subfig}
\usepackage{textcomp}
\usepackage{stfloats}
\usepackage{url}
\usepackage{verbatim}
\usepackage{graphicx}
\usepackage{cite}
\usepackage{times}
\usepackage{epsfig}
\usepackage{amsmath}
\usepackage{amssymb}
\usepackage{booktabs}
\usepackage{utfsym}
\usepackage{pifont}
\usepackage{multirow}
\usepackage{colortbl}
\usepackage{xcolor}
\usepackage{bm}
\usepackage{makecell}
\def\eg{\emph{e.g.}}
\def\ie{\emph{i.e.}}
\def\etal{\emph{et~al.}}
\newcommand{\cmark}{\ding{51}}%
\newcommand{\xmark}{\ding{55}}%

\begin{document}

\title{Online Streaming Video Super-Resolution with Convolutional Look-Up Table}

\author{Guanghao Yin*, Zefan Qu*, Xinyang Jiang, Shan Jiang*, Zhenhua Han, Ningxin Zheng, Xiaohong Liu, Huan Yang, Yuqing Yang, Dongsheng Li, Lili Qiu

\thanks{A preprint version of this research work was put on arXiv. We declare that our manuscript is original and has not been previously published. It is not currently being considered for publication elsewhere.}
\thanks{*Work done during an internship in Microsoft Research Asia.}

\thanks{Guanghao Yin is with the Key Laboratory of Design Intelligence and Digital Creativity of Zhejiang Province, Zhejiang University, Hangzhou, China. Zefan Qu is with the Department of Computer Science and Technology, Tongji University, Shanghai, China. Shan Jiang is with University of
Science and Technology of China, Hefei, China. Xiaohong Liu is with Shanghai Jiao Tong University. Xinyang Jiang, Zhenhua
Han, Ningxin Zheng, Huan Yang, Yuqing Yang, Dongsheng Li
and Lili Qiu are with the Shanghai AI/ML Group, Microsoft Research Asia, Shanghai, China. (Guanghao Yin and Zefan Qu contributed equally to this work.)  (Corresponding author: Xinyang Jiang.)
}
\thanks{E-mail: ygh\_zju@zju.edu.cn, qzf@tongji.edu.cn, jiangshan@ustc.edu.cn, xiaohngliu@sjtu.edu.cn, \{zhenhua.han, ningxin.zheng, huan.yang, yuqing.yang, dongsheng.li, liliqiu, xinyangjiang\}@microsoft.com}}

\markboth{Journal of \LaTeX\ Class Files,~Vol.~14, No.~8, August~2021}%
{Shell \MakeLowercase{\textit{et al.}}: A Sample Article Using IEEEtran.cls for IEEE Journals}


\maketitle

\begin{abstract}
Online video streaming has fundamental limitations on the transmission bandwidth and computational capacity and super-resolution is a promising potential solution. 
  However, applying existing video super-resolution methods to online streaming is non-trivial. 
  Existing video codecs and streaming protocols (\eg, WebRTC) dynamically change the video quality both spatially and temporally, which leads to diverse and dynamic degradations. 
  Furthermore, online streaming has a strict requirement for latency that most existing methods are less applicable. 
  As a result, this paper focuses on the rarely exploited problem setting of online streaming video super resolution.  
  To facilitate the research on this problem, a new benchmark dataset named LDV-WebRTC is constructed based on a real-world online streaming system. 
 Leveraging the new benchmark dataset, we proposed a novel method specifically for online video streaming, which contains a convolution and Look-Up Table (LUT) hybrid model to achieve better performance-latency trade-off. 
 To tackle the changing degradations, we propose a mixture-of-expert-LUT module, where a set of LUT specialized in different degradations are built and adaptively combined to handle different degradations. 
  Experiments show our method achieves 720P video SR around 100 FPS, while significantly outperforms existing LUT-based methods and offers competitive performance compared to efficient CNN-based methods.
\end{abstract}


\section{Introduction}
\label{sec:intro}

\IEEEPARstart{W}{ith} the fast development of network infrastructure, video delivery techniques, and the growing demand of users, video streaming has become the ``killer'' application of the Internet in the past two decades~\cite{lu2018you}. 
Due to users' steep expectations for quality, delivering high-definition (HD) video to end users is important. 
However, the quality of streamed video heavily depends on the network bandwidth between servers and clients. 
Streaming 4K videos require over 40~Mbps bandwidth per user\cite{isobe2020revisiting}, which is difficult to achieve in many areas. With the ever-increasing computational power of client devices and advances in deep learning, super-resolution (SR), which aims to restore high-resolution (HR) frames by adding the missing details from low-resolution (LR) frames, has been considered as a promising direction to reduce the bandwidth requirement of streaming HD videos~\cite{dong2015image,ledig2017photo,ahn2018fast,zhang2018image,caballero2017real,isobe2020revisiting,liu2022learning,chan2022basicvsr++}.

\begin{figure}[t]
  \centering
  \includegraphics[width=8cm]{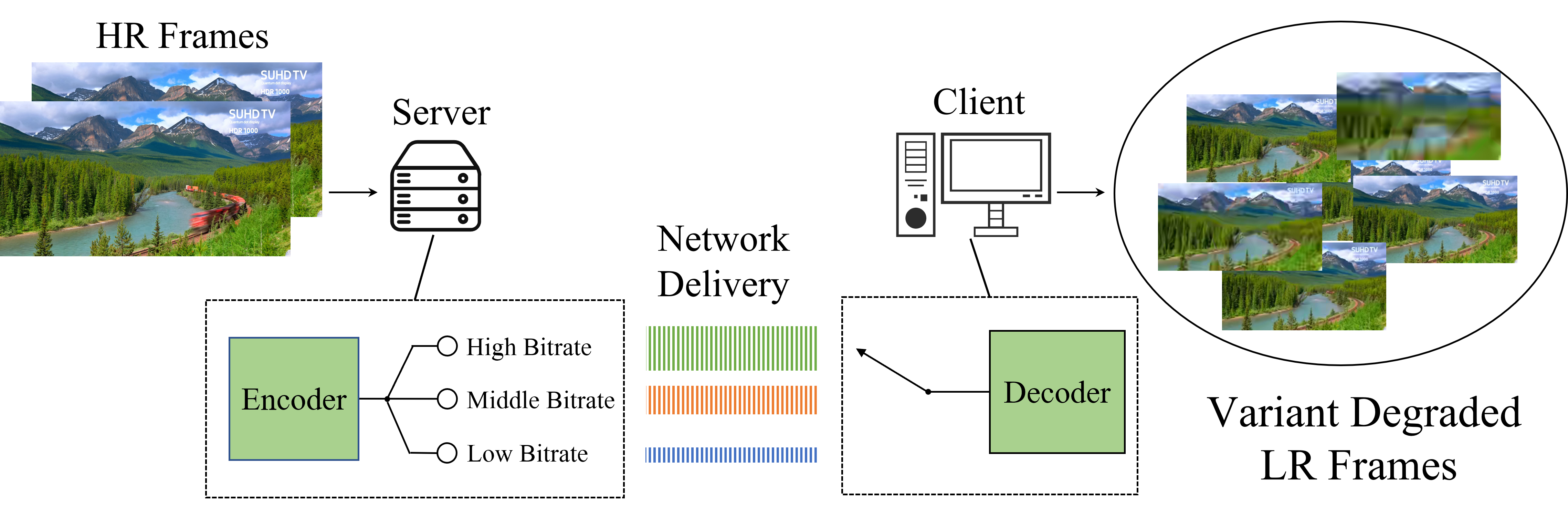}
   \caption{Pipeline of online video streaming. A server uses the Adaptive Bitrate streaming (ABR) to encode frames at multiple bitrates for delivery, and the clients select one of the streams to decode. During this adaptive delivery, the spatial-temporal dynamic degradations are inevitably introduced.}
   \label{fig:streaming}
\end{figure}

While previous works have achieved great progress~\cite{zhang2018image, yang2020learning,ahn2018fast,isobe2020revisiting,liu2022learning,chan2022basicvsr++, qiu2022learning}, existing super-resolution methods still face great challenges in real-world video streaming. Adapting existing video super resolution (VSR) methods to streaming data is non-trivial for two reasons. 
First, it requires SR inference to be fast (\ie, low-latency), especially for online scenarios involving real-time user interaction (\eg, video conference or cloud gaming), where slight latency will significantly harm the user experience. 
Most state-of-the-art VSR methods involve high complexity models and need to cache future frames for super-resolving the current frame, 
 inevitably introducing high latency. 
Second, videos transmitted by streaming system suffer dynamically changing degradations both temporally and spatially.  In temporal domain, in order to adapt the time-varying network conditions, existing video streaming protocols (\eg, WebRTC~\cite{johnston2012webrtc}) adopt bitrate adaptation (ABR), which adaptively changes the quality of the video frame at each time step, As shown in Fig.~\ref{fig:streaming}. 
In spatial domain, the codec used in the streaming system applies different compression configurations to each macro-block within a frame, resulting in degradation variations across different macro-blocks.
Thus, it is essential to develop VSR models capable of dealing with spatial and temporal changing degradations. 

\begin{figure*}[t]
  \centering
  \includegraphics[width=17cm]{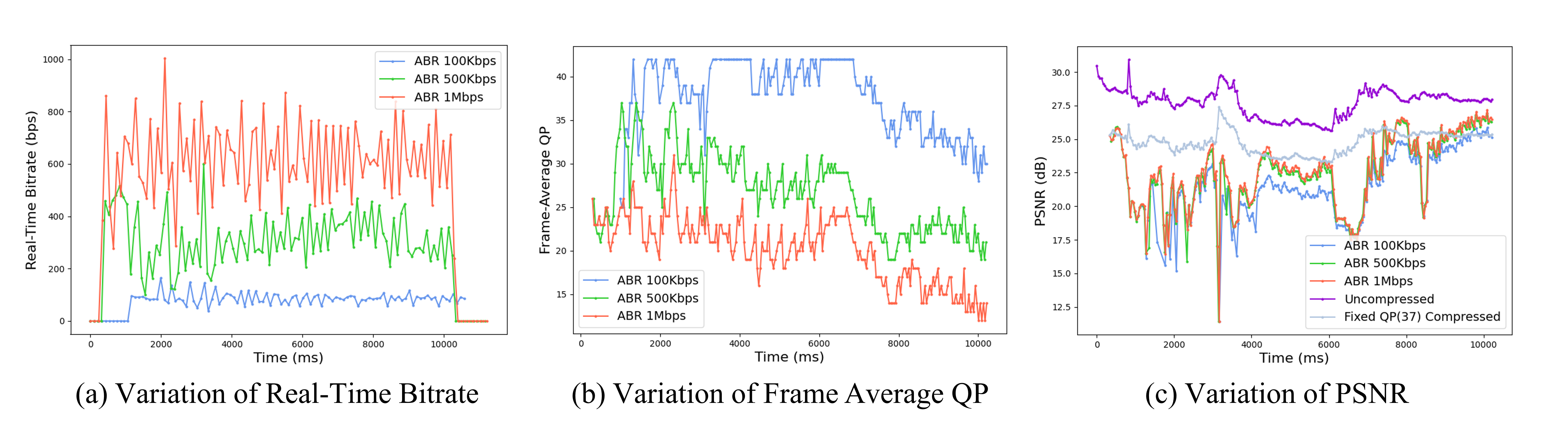}
   \caption{The variations of real-time bitrates, frame-average QP,
   and PSNR of streamed test video $005$ using our LDV-WebRTC testbed under 100Kbps, 500Kbps and 1Mbps bandwidths.} 
   \label{fig:dataset}
\end{figure*}

As a result, this paper focuses on a rarely studied problem setting, which aims at super-revolving videos transmitted by real-world online streaming system, named online streaming VSR. 
Since existing VSR datasets lack videos produced from real-world streaming system,  a new online streaming VSR dataset named LDV-WebRTC is proposed. 
Specifically, we built a real-world online video streaming prototype based on Web Real-Time Communication (WebRTC)~\cite{johnston2012webrtc}, which is used to transmit videos from LDV 2.0 dataset~\cite{yang2022ntire} with ABR under different network conditions.  
As shown in Fig.~\ref{fig:dataset}, compared to un-compressed videos or videos compressed with fixed quantization parameter (QP), the quality and compression configurations of the videos in this dataset varies significantly through time, which further verifies the drastically changing degradations. 
We have developed a benchmark based on this dataset to assess both SR performance and model latency of various VSR baselines. 

Leveraging the proposed dataset, this paper proposes a novel method tackling the aforementioned challenges of online streaming VSR, named ConvLUT. 
To meet the low-latency requirement, we propose a novel hybrid network structure combining convolution and Look-Up Table (LUT) ~\cite{zeng2020learning,wang2021real,liu20224d,jo2021practical,ma2022learning, li2022mulut,zhang2022clut}, resulting in a balanced trade-off between the high inference efficiency of LUT and the strong computational capacity of neural networks.
To tackle the spatial-temporal dynamic degradations, ConvLUT adopts a novel mixture-of-expert-LUT module. It contains a set of expert LUTs specialized in specific degradations, which are built from state-of-the-art (SOTA) SR networks trained on a pool of sampled degradations. The expert LUTs are adaptively combined to address the diverse degradations across different macro-blocks at each time step.  
Furthermore, since the proposed Convolution-LUT hybrid structure is unfriendly to parallel computing accelerators, such as GPU, NPU, and FPGA~\cite{nvidia2018cuda, wang2016performance,khorasani2015efficient}, we further propose an efficient interpolation algorithm to support LUT inference in parallel. 

 In conclusion, our work contributes to threefold:
(1) We propose a rarely studied problem setting of online streaming video super resolution;
(2) We propose the first real-world video streaming SR dataset (LDV-WebRTC) to facilitate the research of online video streaming SR;
(3) A novel conv-LUT hybrid VSR network is proposed which achieves real-time process latency and handles spatial-temporal degradation variation by a mixture of expert LUTs.   

\section{Related Work}
\subsection{Adaptive Online Bitstream}
Adaptive online streaming aims to handle unpredictable bandwidth variations for high-quality video delivery. In online streaming, a server first controls Video Coding Protocols, such as H.264/AVC~\cite{wang2004image} or HEVC~\cite{sullivan2012overview} to encode image frames at multiple bitrates by adjusting QP in both spatial and temporal domains. Then the client uses an ABR algorithm to select suitable video quality and decode the received frames~\cite{yeo2020nemo}. During this adaptive delivery, coding artifacts are inevitably introduced~\cite{chen2020bitstream}. 
Therefore, one of the major challenges for high-quality online streamings is to handle the spatial-temporal dynamic degradations adaptively.

\subsection{Super-Resolution}
Since the pioneering SRCNN~\cite{dong2015image}, deep learning based approaches~\cite{dong2015image,ledig2017photo,zhang2018image,caballero2017real} have exhibited impressive performance in single-image SR (SISR) tasks. By considering the potential dependency
in consecutive frames, various video SR (VSR) models~\cite{ahn2018fast,isobe2020revisiting,chan2021understanding,chan2021basicvsr,liu2022learning,chan2022basicvsr++} have achieved great success, many of which adopts computational intensive modules such as optical flow alignment, deformable convolution and transformer.   
Recently, there has been an increasing interest in efficient SR. For example, CARN~\cite{ahn2018fast} replaces the conventional convolutions with group convolutions, which reduce the parameters of its original big model. VESPCN~\cite{caballero2017real} uses lightweight motion estimation and pixel-shuffle modules to conduct spatial-temporal upscaling. RRN~\cite{isobe2020revisiting} removes the optical flow based alignment, but directly uses hidden states of recurrent proceedings to involve temporal information. 

\subsection{Look-Up Table} 
Look-Up Table is an efficient tool for classic image processing because it can replace complex computations with direct query operations. The pre-defined LUT has been widely used as the template to adjust the pixel distribution in photo editing and camera imaging~\cite{zeng2020learning}. Recent deep models have also extended LUTs to low-level vision tasks~\cite{zeng2020learning,wang2021real,liu20224d,jo2021practical,ma2022learning}. For color enhancement, Zeng \etal~\cite{zeng2020learning} proposes learnable 3D LUT to achieve image-level LUT adaption. Yang \etal~\cite{yang2022adaint} propose a more flexible sampling point allocation to adaptively learn the non-uniform sampling intervals in 3D color space. Liu \etal~\cite{liu20224d} propose a learnable context-aware 4D LUT to achieve content-dependent enhancement. Recently, some new attempts have also been proposed for super-resolution. SR-LUT~\cite{jo2021practical} first use a single 4D LUT to transfer the LR-HR mappings from a pretrained SR model with small receptive field (RF).  SPLUT~\cite{ma2022learning} uses the parallel cascaded LUTs to process the high and low 4-bit components of 8-bit LR images. Meanwhile, the padding aggregations are also applied to enlarge the receptive field of LUT. Nevertheless, the fixed LUT mapping from the simple-designed network structures still limit their performance for dynamic degradations.

\section{Online Streaming VSR and Dataset}
\label{sec:dataset}

Online Streaming VSR is a rarely studied problem setting, and since existing VSR datasets either contain un-compressed raw video frames~\cite{xue2019video,yang2021real,yang2022ntire} or frames compressed with fixed presets~\cite{yang2022ntire}, they do not reflect the spatial-temporal changing degradations of online video streaming. 
Hence, to better tackle real-world challenges and facilitate research on online streaming VSR, we collect a new video SR dataset under real-world online streaming setting, named LDV-WebRTC. 

WebRTC~\cite{johnston2012webrtc} is a real-time communication protocol that is widely used to stream real-time videos to browsers or mobile devices. We build a video streaming prototype based on WebRTC, which uses a server to stream low-resolution videos to a laptop client via a router. The router uses Linux \texttt{tc} to control the network bandwidth between the server and the client to emulate the diverse bandwidth settings of real-world applications. 
We collect all 335 high-resolution videos in the LDV 2.0 dataset~\cite{yang2022ntire}, containing a rich diversity of content scenes whose frame resolution is $960\times 512$. The high-resolution frames are downsampled by $4\times$ in bicubic mode to get $240\times 128$ low-resolution frames, which are encoded via FFmpeg-H.264~\cite{wiegand2003overview} and then transmitted by the servers.  The WebRTC server enables Adaptive Bitrate streaming (ABR) that adjusts encoding quality with QP values in a range of $[0,50]$, according to factors including network bandwidth, encoding latency, and decoding latency. A larger QP leads to worse quality of encoded frames. After receiving an encoding video stream, the client decodes it into a sequence of decoded low-resolution frames with timestamps. The target of video streaming SR is to restore those received LR frames. To involve different network conditions, three representative types of networks are emulated in the router with an average bandwidth of 100kbps, 500kbps, and 1Mbps. Note that, when bandwidth is limited, not all frames are successfully received due to frame drop. Thus we align the decoded frames at the client with the original high-resolution frames using encoding timestamps. In addition to the decoded frames, we also collect the  motion vector priors extracted by the video codec of the streaming system. 

\begin{table}[t]
  \centering
  \renewcommand\tabcolsep{4pt}
  \caption{Statistical Results of our LDV-WebRTC dataset.}
  \begin{tabular}{cccccc}
    \toprule
      Bandwidth & Dataset & \makecell[c]{Bitrate \\ (Kbps)} & QP & \makecell[c]{PSNR\\ (dB)} & \makecell[c]{Frame\\ Number} \\
      \hline
      \multirow{3}*{100Kbps} & Training  & 85.15 & 30.79 & 24.37 & 63792 \\
      ~ & Validation & 84.79 & 29.75 & 23.72 & 4415 \\
      ~ & Test & 82.26 & 29.87 & 23.31 & 5558 \\
      \hline
      \multirow{3}*{500Kbps} & Training  & 334.56 & 20.90 & 25.04 & 74207 \\
      ~ & Validation & 340.05 & 18.80 & 24.68 & 5926 \\
      ~ & Test & 322.05 & 20.42 & 24.27 & 5980 \\
      \hline
      \multirow{3}*{1Mbps} & Training  & 497.22 & 17.73 & 25.21 & 73603 \\
      ~ & Validation & 476.10 & 16.21 & 24.79 & 5939 \\
      ~ & Test & 486.48 & 17.54 & 24.35 & 6078 \\
      \bottomrule
  \end{tabular}
  \label{table:dataset}
  \vspace{-0.4cm}
\end{table}

Fig.~\ref{fig:dataset} illustrates the statistic of real-time bitrates, QP and PSNR of streamed frames using our WebRTC testbed under different network bandwidths. It's clear that the PSNR of real streamed video fluctuated more severely due to the real-time encoding-decoding pipeline and the bitrates variation. The QP values of encoded frames also vary greatly and sometimes even trigger resolution changes (\ie WebRTC's default strategy degrades resolution when the frame-average QP is quite large). Moreover, the frame drop also happens frequently for online streaming. All those observations prove the necessity of building a more realistic dataset that reflects the diverse and time-varying degradation of real-world online video streaming. 

In conclusion, our dataset consists of the aligned LR frames of the client, their original HR versions of the server, and the bitstream priors under 100kbps, 500kbps, and 1Mbps bandwidths. Table~\ref{table:dataset} illustrates the statistical results of average real-time bitrates, QP and PSNR of streamed frames on the training, validation, and test sets, which are collected by our WebRTC testbed under different network bandwidths. When the bandwidth setting decreases from 1Mbps to 100Kbps, the real bitrate decreases and the online streaming system uses lower QP to compress frames, which causes lower image quality. Moreover, the frame drop also happens more frequently when the bandwidth decreases. All those observations reveal the online streaming degradations are dynamic and challenging.

\begin{figure*}[t]
  \centering
   \includegraphics[width=17cm]{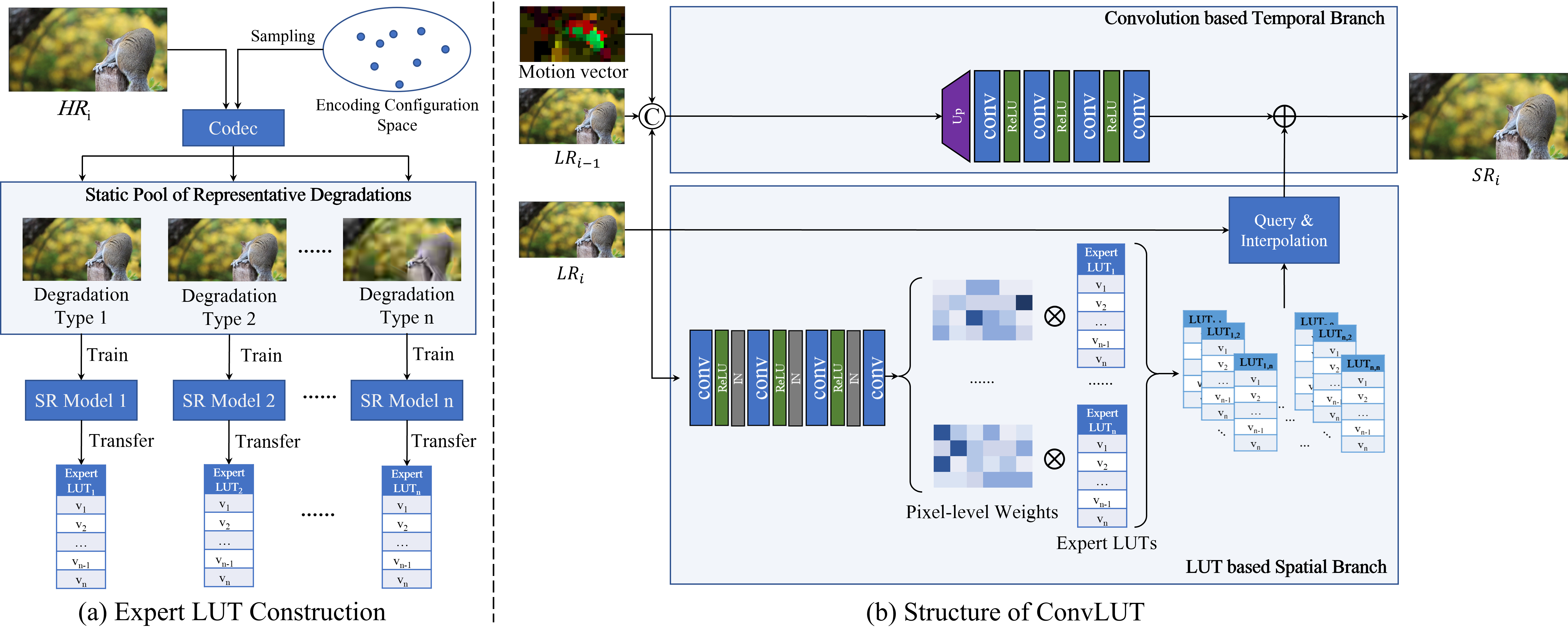}
   \caption{The overall scheme of our proposed ConvLUT. (a) First, a set of $n$ expert LUTs  are built from $n$ SR networks pre-trained on $n$ static representative degradations 
  sampling from the configuration space. (b) Then, the spatial and temporal branches of our ConvLUT conduct adaptive LUT fusion and multiple-frame fusion in parallel, and the outputs of the two branches are added to get the final SR results.}
   \label{fig:framework}
\end{figure*}

\section{Method}
\subsection{ConvLUT Hybrid Network Architecture}
As shown in Fig.~\ref{fig:framework}, we design a hybrid VSR network combining the convolution and LUT for online streaming, which contains two parallel branches. The LUT based spatial branch learns a combination of a group of expert LUTs to handle the dynamic degradation in the spatial dimension, and the convolution based temporal branch further refines the LUT outputs with temporal information from history frames and the priors of codec (i.e. motion vectors). 

\subsection{Mixture of Expert LUT}
\label{sec:spatical}
As explained above, the challenge of online streaming VSR is to handle the dynamic degradations in real-time speed and low latency. 
Despite LUT's fast inference speed and small parameter requirements, existing LUT-based super-resolution (SR) methods ~\cite{jo2021practical,ma2022learning} fail to deliver promising results in online streaming scenarios. This is due to the fact that their LUTs are transferred from a single SR model with a simple network structure, which constrains their ability to adapt to complex degradation. Additionally, these methods infer the LUTs for each pixel solely within a small query patch, which results in limited texture and structural information due to their small receptive fields. This, in turn, creates a significant challenge for online streaming VSR. 
To address it, we propose to create a group of expert LUTs, each of which specializes in a particular type of degradations. Specifically, we propose transferring each expert LUT from a state-of-the-art SR network that has been trained on a particular static degradation.  During inference, the expert LUTs are adaptively fused to handle macro-blocks in each video frame with different degradations. 

In this section, we elaborate on how to construct the expert LUTs (IV.B.1) and how to dynamically fuse the expert LUTs for degraded macro-blocks (IV.B.2). Moreover, we present an efficient LUT query and interpolation method that makes LUT more compatible with parallel acceleration (IV.B.3). 

\subsubsection{Transferring SR Networks to Expert LUTs}
\label{subsec:transfer}
Here, we introduce how to build expert LUTs specialized in different degradations. 
The degradation variation of video streaming systems is primarily caused by changes in the configuration of video compression. In addition to the down-sampling operation, the quantization process used in video compression is the main source of degradation, leading to a loss of details and the introduction of artifacts. 
The combination of various compression parameters leads to an extensive configuration space, making it impractical to explore all possible options. 
As a result, we select a static pool of representative degradations, by sampling compression parameter settings from the configuration space. We then train SR networks on the videos corresponding to each degradation in the pool, obtaining SR models specialized in addressing specific type of degradation. 

As shown in Fig.~\ref{fig:framework}(a), in our experiments, the configuration space is defined by quantization parameter (QP), which is one of the most critical parameters to control quantization process in video compression. 
The higher QP indicates the higher compression ratio and lower video quality. 
We uniformly sample $N$ different QP values from $D_{qp\in[0,m]}$ corresponding to $N$ different degradations $\{D_i\}_{i=0}^N$, and use them to generate $N$ video SR training subsets $\{\mathcal{X}^{D_i}\}_{i=1}^N$. As a result, $N$ SR models $\{f_{sr_i}\}_{i=1}^N$ are trained on those $N$ subsets. It should be noted that the structure of the SISR model has no restrictions and any SOTA methods can be applied for training. 

The pre-trained SR networks are then transferred to expert LUTs specialized in different degradations, which are later dynamically fused to handle degradation variation. Following previous LUT works~\cite{zeng2020learning,jo2021practical,liu20224d,ma2022learning}, each expert LUT stores a mapping between a LR patch and the patch super-resolved by a SR network. 
Specifically, we create a full value permutation of a $2\times2$ patches, which are $256^4$ patches in total ranging from [0,0,0,0] to [255,255,255,255]. 
The SR model $f_{sr_i}$ takes each $2\times 2$ patch as input, and we store all super-resolved $r\times r$ patches into the LUT, which up-samples the left-upper pixel of the low-resolution patch by scale factor $r$. Eventually, when all permutations of input patches are processed by SR model $f_{sr_i}$, we get the transferred $LUT_i$ with the size of $[256,256,256,256,r,r]$. Moreover, we follow SR-LUT~\cite{jo2021practical} to compress $LUT_i$ by uniformly sampling the original LUT with the interval size of 16, resulting in the compressed $LUT_i$ with the size of $[17,17,17,17,r,r]$.

\subsubsection{Adaptive LUT Fusion} 

Given that the degradation of an online streaming video varies both temporally and spatially, it is crucial to obtain a specialized LUT for each macro-block within each frame corresponding to its particular degradation. 
Inspired from the concept of mixture of experts~\cite{jacobs1991adaptive,shazeer2017outrageously,emad2022moesr}, we can create a spatially and temporally variant look-up table by combining expert LUTs with different weight combinations at each pixel position, allowing LUT to effectively adapt to any type of degradation.
Specifically, a lightweight predictor is proposed to output the LUT combination weights for each pixel based on the content of the input frames. The weight predictor, denoted as $f_w$, consists of 4 convolution layers with Instance Normalization~\cite{ulyanov2016instance} and LeakyReLU~\cite{maas2013rectifier} operations. For the input LR frame $X \in \mathbb{R}^{h \times w\times 3}$, $f_w$ outputs a weight tensor $W\in \mathbb{R}^{h \times w\times n}$, where $n$ is the number of expert LUTs. To obtain the SR result of a specific pixel in the frame $X_{i,j,k}$, the weighted LUT used for query is: 
\begin{equation}
  \hat{LUT}_{X_{i,j,k}} = W_{i,j,1} \times LUT_1 + ... +  W_{i,j,n} \times LUT_n.
\end{equation}
It should be noted that the fusion operation is only conducted on the 4D lattice surrounding the input pixel value, not the whole LUT. Moreover, since our weight predictor takes the entire frame as input, the weighted fusion is obtained based on a much larger receptive field, providing more spatial information than previous LUT-based methods~\cite{jo2021practical,ma2022learning}.

\subsubsection{Efficient LUT Interpolation}
\label{subsec:inference}
In order to make the proposed Conv-LUT hybrid
structure more friendly to parallel computing, we introduce an efficient method to query the mixture of expert LUTs. 

Given the 4D input values, the output values of an anchor pixel $X_{i,j,k}$  are generated by querying and interpolating the nearest sampled points in LUT. Specifically, for the input $(x,y,z,u)$, we first conduct the look-up query operation to find its location in the 4D LUT lattice. 
As explained in previous LUT works~\cite{zeng2020learning,jo2021practical,liu20224d,ma2022learning}, the most significant bits (MSBs) of the input pixel value can be used for LUT location, and the least significant bits (LSBs) are used for interpolation. Hence, we separate the 8-bit input pixel values to high 4-bit integers $(h_x,h_y,h_z,h_u)$ and low 4-bit decimals $(l_x,l_y,l_z,l_u)$. Given the high 4-bit integers, we obtain the 4D LUT lattice surrounding the input pixel value, consisting of 16 adjacent sampling points $({h_{x}, h_{x}+1}, {h_{y},h_{y}+1},{h_{z},h_{z}+1},{h_{u},h_{u}+1})$. The low 4-bit decimals $(l_x,l_y,l_z,l_u)$ represent the distance between 4 input pixels $(x,y,z,u)$ to their nearest points.

\begin{figure*}[ht]
  \centering
   \includegraphics[width=17cm]{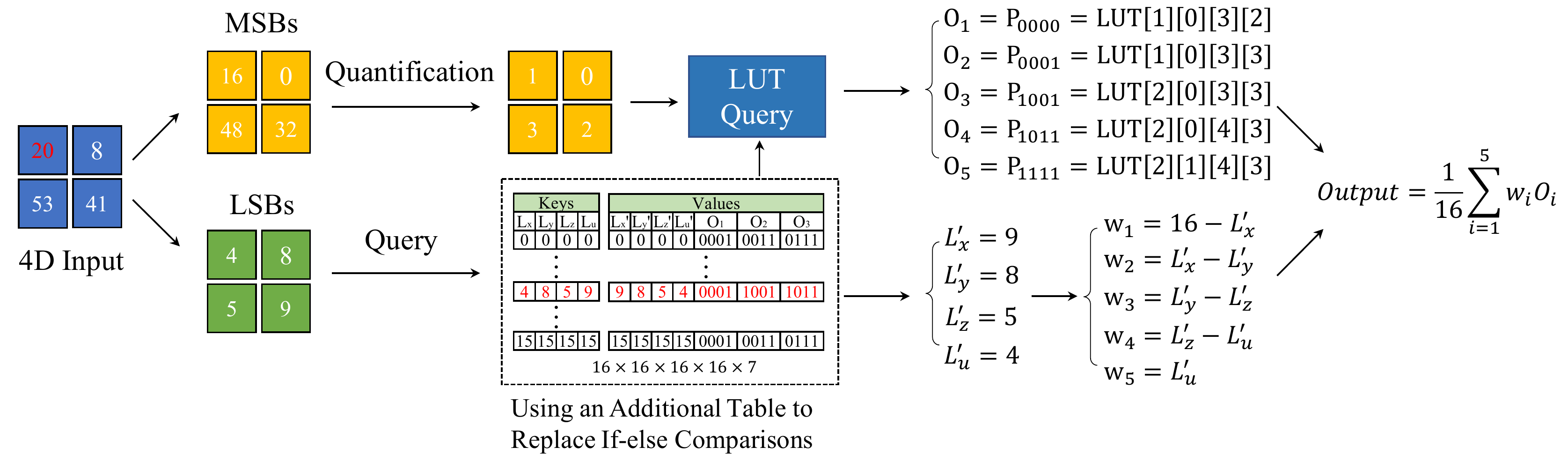}
   \caption{An example of the implementation of our efficient LUT inference. Here, the quantization value is set as 16 to compress LUT. }
   \label{fig:LUT}
\end{figure*}

\begin{table}[t]
\caption{The number of operations of different interpolations. Due to the complicated if-else control flow, tetrahedral interpolation is unfriendly to accelerators like basic CUDA operators. Our interpolation needs fewer operations and can be accelerated by CUDA. }
\small
  \renewcommand\tabcolsep{3.7pt}
  \begin{center}
  
  \begin{tabular}{ccccc}
    \toprule
    Interpolation & Query & Multiplication & \makecell[c]{If-else\\ Control Flow} &\makecell[c]{Parallel\\ Accelerator}\\
    \midrule
    Tetralinear & 16 & 16 & 0 &\cmark \\
    Tetrahedral & 5 & 5 & 24 &\xmark \\
    \textbf{Ours} & 9 & 5 & 0 & \cmark\\
    \bottomrule
  \end{tabular}
  \end{center}
  \label{interpolation}
\end{table}

ConvLUT uses tetrahedral interpolation ~\cite{kasson1995performing}, which needs only 5 multiplications with the values of 5 bounding vertices of 4-simplex geometry. 
However, in practice, finding the 5 vertices among all total 24 neighboring vertices is implemented with 24 control flow instructions, which is unfriendly to parallel accelerators, resulting in a low inference speed. For example, CUDA operators do not support such flow control operation for parallel acceleration.
Hence, we accelerate the tetrahedral interpolation by replacing the complicated control flow with the mapping table query. 
As shown in Table~\ref{table:4D}, the 24 logical statements $(x,y,z,u)$ in tetrahedral interpolation are equivalent to sorting the order of 4 input pixels from small to large. Since all input permutations are countable, we can use an additional table to store the sorted 4 values $(x',y',z',u')$. The indexes of 5 corresponding neighboring vertices $(O_1,O_2,O_3,O_4,O_5)$ can also be store according to the 24 control flow instructions in Table~\ref{table:4D}. It should be noted that the relative indexes of $O_1$ and $O_5$ are kept fixed at 0000 and 1111, and we only need to store the indexes of $(O_2,O_3,O_4)$. Moreover, only the least significant 4-bits determine the weights $(w_1,w_2,w_3,w_4,w_5)$, our additional table only needs to save $16^4$ permutations.

\begin{table}[t]\footnotesize
  \centering
  \renewcommand\tabcolsep{1pt}
  \caption{The 24 control flows of tetrahedral interpolation equivalent for 4D space, also presented in SR-LUT~\cite{jo2021practical}. Since the control flows are unfriendly for parallel accelerators like GPU, we use an additional table to replace the if-else logical operations and uniformly conduct the interpolation.}
  \resizebox{\linewidth}{!}{
  \begin{tabular}{ccccccccc}
    \toprule
      Condition & $w_1$ & $w_2$ & $w_3$ & $w_4$ & $w_5$ & $O_2$ & $O_3$ & $O_4$ \\ 
      \hline
      $L_x{>}L_y{>}L_z{>}L_u$ & $W{-}L_x$ & $L_x{-}L_y$ & $L_y{-}L_z$ & $L_z{-}L_u$ & $L_u$ & $P_{1000}$ & $P_{1100}$ & $P_{1110}$\\
      $L_x{>}L_y{>}L_u{>}L_z$ & $W{-}L_x$ & $L_x{-}L_y$ & $L_y{-}L_u$ & $L_u{-}L_z$ & $L_z$ & $P_{1000}$ & $P_{1100}$ & $P_{1101}$\\
      $L_x{>}L_u{>}L_y{>}L_z$ & $W{-}L_x$ & $L_x{-}L_u$ & $L_u{-}L_y$ & $L_y{-}L_z$ & $L_z$ & $P_{1000}$ & $P_{1001}$ & $P_{1101}$\\
      $L_u{>}L_x{>}L_y{>}L_z$ & $W{-}L_u$ & $L_u{-}L_x$ & $L_x{-}L_y$ & $L_y{-}L_z$ & $L_z$ & $P_{0001}$ & $P_{1001}$ & $P_{1101}$\\
      $L_x{>}L_z{>}L_y{>}L_u$ & $W{-}L_x$ & $L_x{-}L_z$ & $L_z{-}L_y$ & $L_y{-}L_u$ & $L_u$ & $P_{1000}$ & $P_{1010}$ & $P_{1110}$\\
      $L_x{>}L_z{>}L_u{>}L_y$ & $W{-}L_x$ & $L_x{-}L_z$ & $L_z{-}L_u$ & $L_u{-}L_y$ & $L_y$ & $P_{1000}$ & $P_{1010}$ & $P_{1011}$\\
      $L_x{>}L_u{>}L_z{>}L_y$ & $W{-}L_x$ & $L_x{-}L_u$ & $L_u{-}L_z$ & $L_z{-}L_y$ & $L_y$ & $P_{1000}$ & $P_{1001}$ & $P_{1011}$\\
      $L_u{>}L_x{>}L_z{>}L_y$ & $W{-}L_u$ & $L_u{-}L_x$ & $L_x{-}L_z$ & $L_z{-}L_y$ & $L_y$ & $P_{0001}$ & $P_{1001}$ & $P_{1011}$\\
      $L_z{>}L_x{>}L_y{>}L_u$ & $W{-}L_z$ & $L_z{-}L_x$ & $L_x{-}L_y$ & $L_y{-}L_u$ & $L_u$ & $P_{0010}$ & $P_{1010}$ & $P_{1110}$\\
      $L_z{>}L_x{>}L_u{>}L_y$ & $W{-}L_z$ & $L_z{-}L_x$ & $L_x{-}L_u$ & $L_u{-}L_y$ & $L_y$ & $P_{0010}$ & $P_{1010}$ & $P_{1011}$\\
      $L_z{>}L_u{>}L_x{>}L_y$ & $W{-}L_z$ & $L_z{-}L_u$ & $L_u{-}L_x$ & $L_x{-}L_y$ & $L_y$ & $P_{0010}$ & $P_{0011}$ & $P_{1011}$\\
      $L_u{>}L_z{>}L_x{>}L_y$ & $W{-}L_u$ & $L_u{-}L_z$ & $L_z{-}L_x$ & $L_x{-}L_y$ & $L_y$ & $P_{0001}$ & $P_{0011}$ & $P_{1011}$\\
      $L_y{>}L_x{>}L_z{>}L_u$ & $W{-}L_y$ & $L_y{-}L_x$ & $L_x{-}L_z$ & $L_z{-}L_u$ & $L_u$ & $P_{0100}$ & $P_{1100}$ & $P_{1110}$\\
      $L_y{>}L_x{>}L_u{>}L_z$ & $W{-}L_y$ & $L_y{-}L_x$ & $L_x{-}L_u$ & $L_u{-}L_z$ & $L_z$ & $P_{0100}$ & $P_{1100}$ & $P_{1101}$\\
      $L_y{>}L_u{>}L_x{>}L_z$ & $W{-}L_y$ & $L_y{-}L_u$ & $L_u{-}L_x$ & $L_x{-}L_z$ & $L_z$ & $P_{0100}$ & $P_{0101}$ & $P_{1101}$\\
      $L_u{>}L_y{>}L_x{>}L_z$ & $W{-}L_u$ & $L_u{-}L_y$ & $L_y{-}L_x$ & $L_x{-}L_z$ & $L_z$ & $P_{0001}$ & $P_{0101}$ & $P_{1101}$\\
      $L_y{>}L_z{>}L_x{>}L_u$ & $W{-}L_y$ & $L_y{-}L_z$ & $L_z{-}L_x$ & $L_x{-}L_u$ & $L_u$ & $P_{0100}$ & $P_{0110}$ & $P_{1110}$\\
      $L_y{>}L_z{>}L_u{>}L_x$ & $W{-}L_y$ & $L_y{-}L_z$ & $L_z{-}L_u$ & $L_u{-}L_x$ & $L_x$ & $P_{0100}$ & $P_{0110}$ & $P_{0111}$\\
      $L_y{>}L_u{>}L_z{>}L_x$ & $W{-}L_y$ & $L_y{-}L_u$ & $L_u{-}L_z$ & $L_z{-}L_x$ & $L_x$ & $P_{0100}$ & $P_{0101}$ & $P_{0111}$\\
      $L_u{>}L_y{>}L_z{>}L_x$ & $W{-}L_u$ & $L_u{-}L_y$ & $L_y{-}L_z$ & $L_z{-}L_x$ & $L_x$ & $P_{0001}$ & $P_{0101}$ & $P_{0111}$\\
      $L_z{>}L_y{>}L_x{>}L_u$ & $W{-}L_z$ & $L_z{-}L_y$ & $L_y{-}L_x$ & $L_x{-}L_u$ & $L_u$ & $P_{0010}$ & $P_{0110}$ & $P_{1110}$\\
      $L_z{>}L_y{>}L_u{>}L_x$ & $W{-}L_z$ & $L_z{-}L_y$ & $L_y{-}L_u$ & $L_u{-}L_x$ & $L_x$ & $P_{0010}$ & $P_{0110}$ & $P_{0111}$\\
      $L_z{>}L_u{>}L_y{>}L_x$ & $W{-}L_z$ & $L_z{-}L_u$ & $L_u{-}L_y$ & $L_y{-}L_x$ & $L_x$ & $P_{0010}$ & $P_{0011}$ & $P_{0111}$\\
      $else$ & $W{-}L_u$ & $L_u{-}L_z$ & $L_z{-}L_y$ & $L_y{-}L_x$ & $L_x$ & $P_{0001}$ & $P_{0011}$ & $P_{0111}$\\
  \bottomrule
  \end{tabular}
  }
  \vspace{-0.4cm}
  \label{table:4D}
\end{table}

An example of our efficient LUT inference is presented in Fig.~\ref{fig:LUT}. For the input $(x,y,z,u)$, we first separate the most significant bits (MSBs) $(H_x,H_y,H_z,H_u)$ and the least significant bit (LSBs) $(L_x,L_y,L_z,L_u)$. The MSBs of the input pixel value can be used for LUT location and least significant bits (LSBs) are used for interpolation. Specifically, we separate the 8bit input pixel values to high 4bit integers $(H_x,H_y,H_z,H_u)$ as:
\begin{equation}
  H_{x} =\left\lfloor \frac{x}{16}\right\rfloor,H_{y}=\left\lfloor \frac{y}{16}\right\rfloor,H_{z}=\left\lfloor \frac{z}{16}\right\rfloor,H_{u}=\left\lfloor \frac{u}{16}\right\rfloor,
  \label{MSBs}
\end{equation}
and low 4bit decimals $(L_x,L_y,L_z,L_u)$ as:
\begin{equation}
  \begin{split}
    L_{x} =x-H_{x}\times W, L_{y} =y-H_{y}\times W,\\L_{z} =z-H_{z}\times W,L_{u} =u-H_{u}\times W,
  \end{split}
  \label{LSBs}
\end{equation}
where $\left\lfloor \cdot\right\rfloor$ is the floor function, and $W$ represents the quantization value to compress the LUT, which is set as 16 in our paper. Then, we query the pre-defined additional table to get the sorted LSBs $(L_x',L_y',L_z',L_u')$ and the binary indexes of $(O_2,O_3,O_4)$. When we get the sorted LSBs $(L_x',L_y',L_z',L_u')$, the calculation of interpolation weights $(w_1,w_2,w_3,w_4,w_5)$ dose not need the if-else judgement in Table~\ref{table:4D}, but can be uniformly defined as: 
\begin{equation}
  \begin{aligned}
    w_1 = W - &L_x', w_2 = L_x' - L_y', w_3 = L_y' - L_z',\\ &w_4 = L_z' - L_u', w_5 = L_u'.
  \end{aligned}
\end{equation}
When we get the binary indexes of $(O_2,O_3,O_4)$, the values of $(O_1,O_2,O_3,O_4,O_5)$ can be accessed by using their binary indexes and MSBs $(H_x,H_y,H_z,H_u)$ to conduct LUT query. For example, if the index of $O_2$ is 0001, the value of $O_2$ is $P_{0001} = LUT[H_x][H_y][H_z][H_u+1]$. Finally, the output of weighted interpolation are calculated as: 
\begin{equation}
  Output = \frac{1}{W}\sum_{i=1}^5 w_i*O_i.
\end{equation}
During the LUT interpolation, the weighted combination of $N$ LUT bases can also be conducted in parallel.

Since only the sorted LSBs determine the interpolation weights, the size of the mapping table can be efficiently compressed to $[16,16,16,16,7]$ by only storing 4-bit 4D-LSB permutations instead of $256^4$ 8-bit 4D-pixel permutations. Moreover, the relative index of $O_0$ and $O_4$ are kept fixed at 0000 and 1111, and we only need to store the index of $(O_1,O_2,O_3)$. Once the order table is pre-defined, the comparison and flow control operation can be replaced by query, and the LUT inference can be accelerated in parallel. 

Therefore, our accelerated LUT inference can be defined as three steps: (1) For the input $(x,y,z,u)$, we first separate the MSBs $(h_x,h_y,h_z,h_u)$ and LSBs $(l_x,l_y,l_z,l_u)$; (2) we query the pre-defined order table to get the sorted LSBs  $(l_x',l_y',l_z',l_u')$ and the binary index of $(O_1,O_2,O_3)$; (3) we conduct the unified tetrahedral interpolation with the additional order table. As shown in Table~\ref{interpolation}, the number of operations of our accelerated interpolation is smaller and can be easily deployed to accelerators. In our work, we use CUDA accelerator to conduct parallel computation.

\subsection{Temporal Branch for Multi-Frame Processing}
To fully utilize the temporal information, our proposed model contains a temporal branch responsible for refining the result of the LUT branch with history frames and object motions. 
While the LUT branch utilizes a look-up table to handle spatial degradation, the temporal branch uses convolution networks with better computational capacity to handle more complicated temporal and motion related information. 

To achieve fast speed and low latency for online streaming, commonly used optical flow alignment, deformable convolution and transformer~\cite{ahn2018fast,isobe2020revisiting,liu2022learning,chan2022basicvsr++} are not suitable for our task. 
Moreover, due to the latency restriction of online VSR, the future frames cannot be utilized.

As shown in Fig.~\ref{fig:framework}(b), our temporal branch only uses 4 conventional layers with LeakyReLU~\cite{maas2013rectifier} to fuse the previous and current frames. To incorporate motion-related information, we leverage a readily available video streaming prior: motion vectors. Similar to optical flow, these vectors provide a coarse approximation of patch-level correspondence and alignment between two frames. However, unlike optical flow, they require no extra calculation since they are part of the streaming system's prior knowledge. To avoid additional computational cost, we simply add the motion vector between two frames as an additional feature map.

\begin{table*}[t]
 \caption{$\times 4$ SR model comparisons on LDV-WebRTC testsets under 100Kbps, 500Kbps, and 1Mbps. The latency levels required by different methods are sorted from high to low. Size denotes the storage space or the parameter number of each model. The row highlighted in \textcolor{gray}{GRAY} means the SR method has unbearable high latency, and thus cannot be applied for online streaming. For online practical SR methods, best and second best results are highlighted in \textcolor{red}{RED} and \textcolor{blue}{BLUE}. Runtime is measured with 2080Ti GPU for generating $1280\times 720$ results. *: The LUT-based methods are accelerated by our interpolation. $\dagger$: The storage space of the LUT-based method.}
\small
  \begin{center}
  \renewcommand\tabcolsep{2pt} 
  \begin{tabular}{c|c|c|c|c|c|c|c|cc}
    \toprule
      Latency & \multirow{2}*{Model} & 100Kbps &500Kbps &1Mbps & \multirow{2}*{Size} & \multirow{2}*{Runtime} & \multirow{2}*{FPS}&Video&Gaming\\ 
       Level&  & PSNR / SSIM & PSNR / SSIM & PSNR / SSIM & ~ & ~ & ~ & (30FPS) & (60FPS) \\ 
      \hline
      \multirow{4}*{High} 
      ~ & \cellcolor{gray!40}BasicVSR++~\cite{chan2022basicvsr++}& \cellcolor{gray!40}23.95 / 0.6210&\cellcolor{gray!40}24.70 / 0.6754&\cellcolor{gray!40}25.11 / 0.6997&\cellcolor{gray!40}9.54M&\cellcolor{gray!40}418.5ms&\cellcolor{gray!40}2.34& \multirow{4}*{\xmark}& \multirow{4}*{\xmark}\\ 
      ~ & \cellcolor{gray!40}TTVSR~\cite{liu2022learning}&\cellcolor{gray!40}23.87 / 0.6246&\cellcolor{gray!40}24.73 / 0.6742&\cellcolor{gray!40}25.24 / 0.7012&\cellcolor{gray!40}6.72M&\cellcolor{gray!40}244.3ms&\cellcolor{gray!40}4.09&~ & ~ \\ 
      ~ &\cellcolor{gray!40}RCAN~\cite{zhang2018image}&\cellcolor{gray!40}23.80 / 0.6165&\cellcolor{gray!40}24.64 / 0.6729&\cellcolor{gray!40}24.85 / 0.6960&\cellcolor{gray!40}15.6M&\cellcolor{gray!40}205.4ms&\cellcolor{gray!40}4.89&~ & ~ \\ 
      ~ &\cellcolor{gray!40}SRResNet~\cite{ledig2017photo}&\cellcolor{gray!40}23.76 / 0.6141&\cellcolor{gray!40}24.53 / 0.6663&\cellcolor{gray!40}24.62 / 0.6878&\cellcolor{gray!40}1.52M&\cellcolor{gray!40}75.80ms& \cellcolor{gray!40}13.19&~ & ~ \\ 
      \hline
      \multirow{4}*{Middle} & RRN~\cite{isobe2020revisiting} & 23.48 / 0.6121 & 24.48 / 0.6521 & \textcolor{blue}{24.84} / 0.6811 & 3.36M & 27.00ms & 37.03&\multirow{4}*{\cmark}&\multirow{4}*{\xmark}\\ 
      ~ & CARN~\cite{ahn2018fast} & 23.77 / 0.6137 & 24.50 / 0.6565 & 24.74 / \textcolor{red}{0.6927} & 1.59M & 25.30ms & 39.53&~ & ~\\ 
      ~ & VESPCN~\cite{caballero2017real} & 23.25 / 0.6063 & 24.34 / 0.6427 & 24.50 / 0.6774 & 0.88M & 22.84ms & 43.78&~ & ~\\ 
      ~ & SPLUT*~\cite{ma2022learning} & 23.34 / 0.6097 & 24.41 / 0.6333 & 24.55 / 0.6764 & 18.12MB$\dagger$ & 21.81ms & 45.85&~ & ~\\ 
      \hline
      \multirow{6}*{Low} & BI  & 23.31 / 0.6031 & 24.27 / 0.6274 & 24.35 / 0.6724 & - & - & -&\multirow{6}*{\cmark}&\multirow{6}*{\cmark}\\ 
      ~ & PAN~\cite{zhao2020efficient} & 23.55 / 0.6123 & 24.50 / 0.6332 & 24.66 / 0.6742 & 0.27M & 16.12ms & 62.50&~ & ~\\ 
      ~ & SR-LUT*~\cite{jo2021practical} & 23.45 / 0.5901 & 24.32 / 0.6198 & 24.42 / 0.6451 & \textcolor{red}{1.27MB$\dagger$} & \textcolor{blue}{11.60ms} & \textcolor{blue}{79.37}&~ & ~\\ 
      \cline{2-8}
      ~ & ConvLUT-CARN* & 23.73 / 0.6123 & 24.46 / 0.6540 & 24.80 / 0.6803 & \multirow{3}*{\textcolor{blue}{8.65MB$\dagger$}}& \multirow{3}*{\textcolor{red}{10.23ms}} & \multirow{3}*{\textcolor{red}{97.75}}&~ & ~\\ 
      ~ & ConvLUT-SRResNet* & \textcolor{blue}{23.78} / \textcolor{blue}{0.6140} & \textcolor{red}{24.54} / \textcolor{blue}{0.6558} & \textcolor{red}{24.87} / \textcolor{blue}{0.6822} & ~& ~ & ~&~ & ~\\ 
      ~ & ConvLUT-RCAN* & \textcolor{red}{23.82} / \textcolor{red}{0.6163} & \textcolor{blue}{24.52} / \textcolor{red}{0.6576} & 24.82 / 0.6816 & ~& ~ & ~&~ & ~\\ 
      \bottomrule
  \end{tabular}
  \end{center}
  \label{table:SOTA}
\end{table*}

\section{Experiment}
\subsection{Experimental Setting}
\noindent\textbf{Datasets.}
The experiments are conducted on the proposed real-world online streaming VSR dataset, LDV-WebRTC. We focus on the scale factor $r=4$. To assess the model ability to deal with degradations under different bandwidths, we use the LR-HR pairs under 1Mbps for training, and evaluate SR models on 100Kbps, 500Kbps, and 1Mbps testsets respectively. 

\noindent\textbf{Evaluation Metrics.}
We evaluate the SR performance for online streaming from three perspectives: the number of model parameters, runtime, and the distortion quality of the generated results. Specifically, Peak Signal-to-Noise Ratio (PSNR) and Structural Similarity Index (SSIM)~\cite{wang2004image} are adopted for evaluation. To compare the running speed, we measure and report the runtime of super-resolving 320 × 180 LR images on one NVIDIA RTX 2080Ti GPU.

\noindent\textbf{Implementation Details.}
As explained in Sec.~\ref{subsec:transfer}, we uniformly select 6 QP values $(QP=0,10,20,30,40,50)$ and use these QP values to encode 6 degraded video subsets $\{D_1, ..., D_6\}$. Three state-of-the-art SR models,  SRResNet~\cite{ledig2017photo}, CARN~\cite{ahn2018fast}, and RCAN~\cite{zhang2018image} are trained on the 6 datasets and transferred to 3 groups of expert LUTs. Finally, ConvLUT is trained with the expert LUTs, resulting in three models, denoted as ConvLUT-SRResNet, ConvLUT-CARN, and ConvLUT-RCAN.

The number of channels of spatial and temporal branches is set to 64. The number of output channels of pixel-level weight predictor is set to 6, matching the number of expert LUTs. In training configurations, the image patch is randomly cropped with the size of $48\times48$, and the batch size is set to 16. The whole ConvLUT is jointly trained by imposing Charbonnier  loss on the final SR outputs. We use Adam optimizer with $\beta_1=0.9$ and $\beta_2=0.999$ to update model parameters. The initial learning rate is $10^{-4}$. We conducted the model training with NVIDIA Tesla V100 GPUs.

\begin{figure*}[t]
  \centering
   \includegraphics[width=17.5cm]{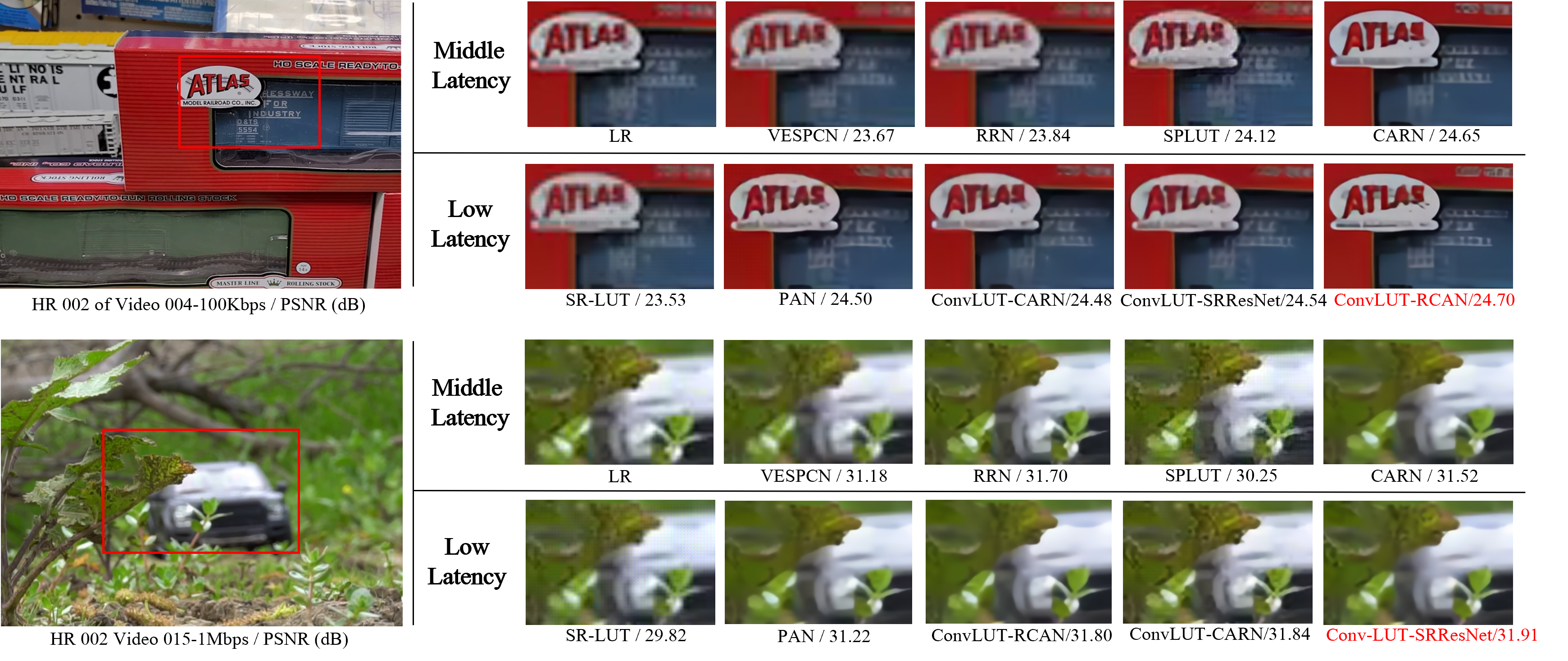}
   \caption{SR perceptual results (×4) of images selected from 100Kbps and 1Mbps testsets. Best results are highlighted in \textcolor{red}{red}.}
   \label{fig:visual}
\end{figure*}

\subsection{Experiments on LDV-WebRTC Dataset}
\label{sec:SOTA}
We compare our method with various SOTAs. Among these methods, SRResNet~\cite{ledig2017photo}, CARN~\cite{ahn2018fast}, and RCAN~\cite{zhang2018image} are the base networks for expert LUT construction. PAN~\cite{zhao2020efficient} is a widely used single-image SR structure with low model complexity and fast inference speed. We also compare with two fast VSR methods VESPCN~\cite{caballero2017real} and RRN~\cite{isobe2020revisiting}. Due to low latency requirement in online streaming scenario, it is infeasible to cache future frames for VSR models. As a result, we modify VESPCN by removing the next frame branch and only extract the spatial-temporal information with the previous and current frames.
BasicVSR++~\cite{chan2022basicvsr++} and TTVSR~\cite{liu2022learning} are much larger bi-directional VSR models. Although it is infeasible to apply them to online scenarios, we still involve them as the performance upper bound. Moreover, we further compare two  LUT-based SISR models, SR-LUT~\cite{jo2021practical} and SPLUT~\cite{ma2022learning}. Note that the GPU inference speed of the current SR-LUT implementation is quite slow due to the large portion of if-else control flow operations and serial for-loop processing for each pixel. For a fair comparison, we also accelerate those LUT-based models with our efficient interpolation method in parallel, as explained in Sec.~\ref{subsec:inference}. We used the open-source codes provided by the authors to implement the compared methods. All methods use the same train-test set partition. 

All the evaluation results are reported in Table~\ref{table:SOTA}. 
Due to the strict requirement of latency for online streaming, we further measure runtime and FPS to evaluate the model efficiency. 
Based on FPS, We categorize the compared SR methods into three types. We categorize methods with FPS lower than 30 as High Latency methods, which are hard to support online streaming applications. Low Latency group refers to methods with FPS higher than 60, fast enough to support high-frame-rate gaming. The rest methods in the range of 30 FPS and 60 FPS are marked as Middle Latency. Benefited from spatial-temporal feature extraction and deep network structure, VSR methods BasicVSR++~\cite{chan2022basicvsr++} and TTVSR~\cite{liu2022learning}, as well as SISR models RCAN~\cite{zhang2018image} and SRResNet~\cite{ledig2017photo} achieve high PSNR/SSIM performance, in exchange of very high latency impractical to online streaming. Compared with methods in the Middle Latency group, our model outperforms them with a marginal improvement in terms of PSNR and SSIM, and in the meantime achieves a much better latency and FPS.  In low latency scenarios, our method can significantly outperform other lightweight models such as  PAN~\cite{zhao2020efficient} and LUT-based SR-LUT~\cite{jo2021practical} in terms of PSNR/SSIM values and meanwhile achieves the lowest latency. The comparisons between our models using 3 types of expert LUTs can also give some interesting findings. When the quality of LR frames decreases (\ie 100Kbps), the expert LUTs transferred from better SR structures (\ie RCAN) produce better results. Moreover, by comparing model sizes, we can see that our method only brings a linear increase in storage costs because of multiple expert LUTs. We believe the model size of our ConvLUT is acceptable for current devices. All those results verify the effectiveness of our ConvLUT in the online streaming scenario. 

\begin{figure*}[t]
  \centering
   \includegraphics[width=17.5cm,trim=250 80 250 60,clip]{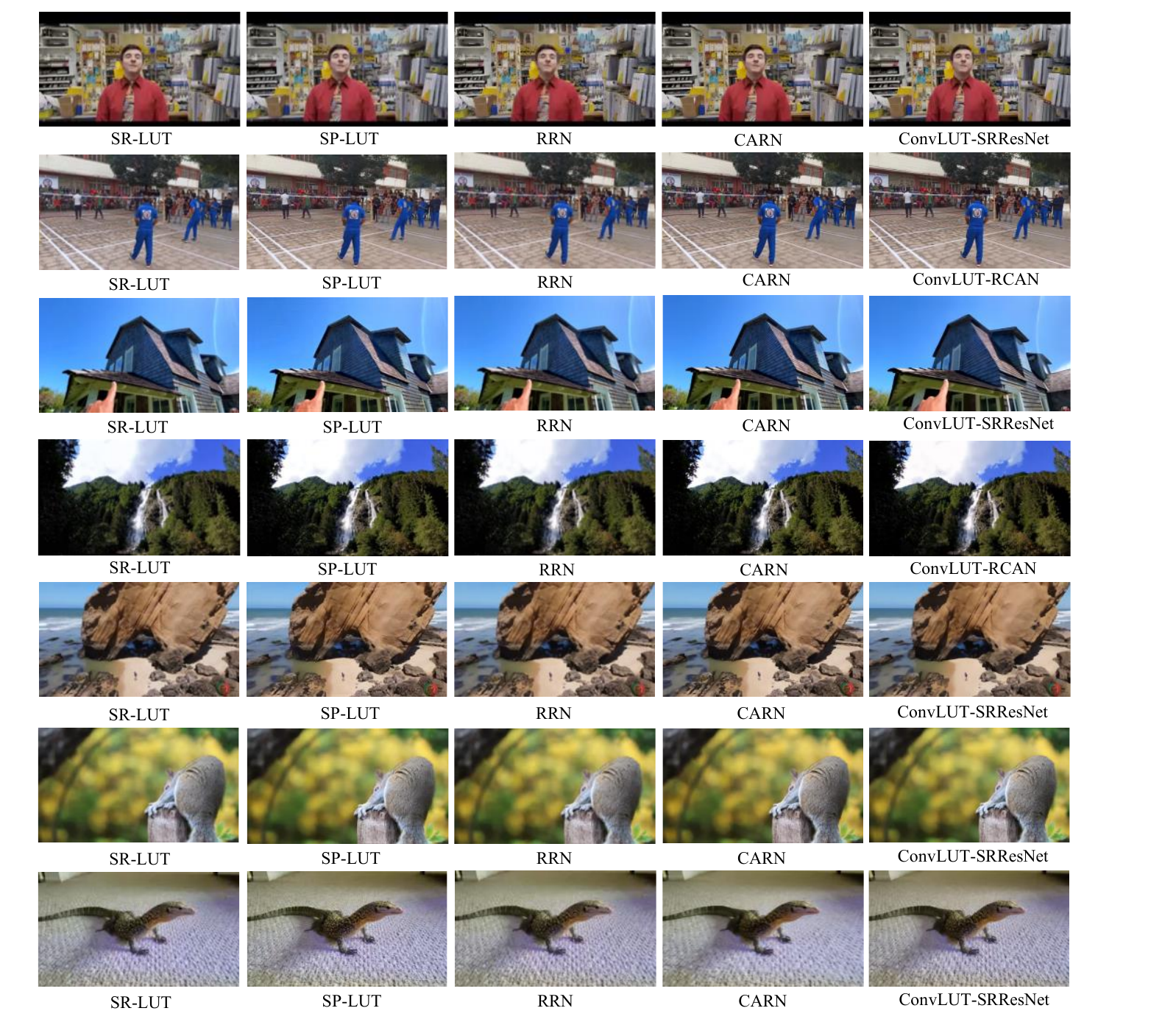}
   \caption{More qualitative comparisons of SR-LUT~\cite{jo2021practical}, SP-LUT~\cite{ma2022learning}, RRN~\cite{isobe2020revisiting}, and CARN~\cite{ahn2018fast} on our LDV-WebRTC testset. The frames listed are from video 007, 013, 010, 011, 012, 001 and 002.}
   \label{fig:morevisual}
\end{figure*}

Visual comparisons are shown in Fig.~\ref{fig:visual}. Here, we choose SR methods relatively practical for streaming in the Middle and Low Latency groups for comparisons. It can be seen that previous LUT-based methods, including SR-LUT~\cite{jo2021practical} and SPLUT~\cite{ma2022learning}, fail to present natural details and produce more artifacts like blocking effect. Limited by fewer conventional layers and network parameters, the lightweight CNN-based SR models, such as PAN~\cite{zhao2020efficient} and VESPCN~\cite{caballero2017real}, generate results with fewer high-frequency details. And our models produce fewer artifacts than previous LUT-based methods and present a similar level of sharpness to CARN~\cite{ahn2018fast} and RRN~\cite{isobe2020revisiting} at faster speed. In addition to the detailed comparison in Fig. \ref{fig:visual}, we also demonstrate the advantages of our approach over some typical rapid super-resolution networks with more samples in Fig. \ref{fig:morevisual}.

\begin{table}[t]
\caption{Ablation studies of the components of ConvLUT on 1Mbps testset. The expert LUTs are transferred from RCAN~\cite{zhang2018image}. Each component brings improvements in terms of PSNR.
}
\small
  \renewcommand\tabcolsep{3pt} 
  \begin{center}
  
  \begin{tabular}{l|c|c|c|c}
    \toprule
    ~& (A) & (B) & (C) & \textbf{Our}\\
    \midrule
    Single-LUT&\cmark&~&~&~\\
    Multi-LUTs Adaptive Fusion&~&\cmark&\cmark&\cmark\\
    Previous Frame Information&~&~&\cmark&\cmark\\
    Motion Vector information&~&~&~&\cmark\\
    \midrule
    Size (MB)&1.27&8.52&8.64&8.65\\
    Runtime (ms)&2.84&4.07&10.21&10.23\\
    PSNR (dB)&24.21&24.47&24.80&\textbf{24.84}\\
    \bottomrule
  \end{tabular}
  \end{center}
\label{table:component}

\end{table}

\subsection{Experiments on Static Degradation}
The expert LUTs of our method are determined by the structure of SISR model and the corresponding training degradations of those SR models. 
Here, we evaluate the effectiveness of the fusion of expert LUTs with different SR network structures, which are trained with the same static degradation. Specifically, we follow the NTIRE 2022 challenge~\cite{yang2022ntire} to conduct experiments on the static $QP=37$ compression degradation. Three SISR model SRResNet~\cite{ledig2017photo}, CARN~\cite{ahn2018fast}, and RCAN~\cite{zhang2018image} are used as the base networks for LUT construction, and our model is denoted as ConvLUT-RCAN+SRResNet+CARN. All those 3 SR networks are trained with the same training set and three corresponding transferred LUTs are grouped as the bases for LUT fusion. And we follow the NTIRE 2022 challenge~\cite{yang2022ntire} to train and test SR models on LDV 2.0 dataset. 

The SOTA comparisons are presented in Table~\ref{table:NTIRE}. Except the 3 mentioned SR models above, we further add 3 novel methods presented in NTIRE 2022 Report~\cite{yang2022ntire}. Although cannot outperform the deep VSR methods, such as the 1st winner model GY-Lab, our model still outperforms both LUT-based models and three SR base networks. Those results proves that the combination of LUT bases can also efficiently fuse the capabilities of different SR structures. In practice, the LUT fusion can be considered in both the training degradation and the SR structure.

\begin{table}[t]
  \small
  \centering
  \renewcommand\tabcolsep{1pt} 
  \caption{$\times 4$ SR model comparisons of NTIRE 2022 Challenge Track3 on Super-Resolution and Quality Enhancement of Compressed Video. The results from the 1st row to 3rd row are directly extracted from NTIRE 2022 Report~\cite{yang2022ntire}. *: The LUT-based methods are accelerated by our interpolation.}
  \resizebox{\linewidth}{!}{
  \begin{tabular}{lccc}
    \toprule
      Model & PSNR & Runtime & Hardware \\
      \hline
      GY-Lab (BasicVSR++~\cite{chan2022basicvsr++} based) 1st~\cite{yang2022ntire} & 24.23 & 11.5s & V100 \\
      AVRT (VSRTransformer~\cite{cao2021video} based) 9th~\cite{yang2022ntire} & 23.52 & 2s & A100 \\
      Modern\_SR (EDVR~\cite{wang2019edvr} based) 12th~\cite{yang2022ntire} & 23.03 & 860ms & 3080 \\
      \hline
      RCAN~\cite{zhang2018image} & 22.96 & 205.4ms & 2080ti \\
      SRResNet~\cite{ledig2017photo} & 22.93 & 75.80ms & 2080ti \\
      CARN~\cite{ahn2018fast} & 22.89 & 25.30ms & 2080ti \\
      SP-LUT*~\cite{ma2022learning} & 22.87 & 21.81ms & 2080ti \\
      SR-LUT*~\cite{jo2021practical} & 22.61 & 11.60ms & 2080ti \\
      ConvLUT-RCAN+SRResNet+CARN*& 23.32 & 10.23ms & 2080ti \\
      \bottomrule
  \end{tabular}
  }
  \label{table:NTIRE}
\end{table}

\subsection{Ablation Analysis}
\noindent\textbf{Effectiveness of Different Modules.}
Table~\ref{table:component} shows the effectiveness of different modules in ConvLUT. 
The baseline model (A),  where only one LUT is applied, is quite similar to SR-LUT~\cite{jo2021practical}, and it produces poor PSNR result, which proves the small receptive field and single neural networks cannot handle the dynamic degradations of online streaming. 
 By adding adaptive LUT fusion, the PSNR has increased by 0.26dB, showing that the pixel-level weight predictor effectively fuses  the expert LUTs to adaptively handle dynamic degradations. 
 The weight predictor cost more storage and computation, but the 8.52MB increase should be acceptable in practice. 
 After adding temporal information from previous frames and the motion vector priors, the performances of our model are further improved, at the cost of lower inference speed. The commonly used temporal fusion modules such as optical flow estimation or deformable convolution are too slow to be applied to online streaming, and hence we adopt a lightweight structure to process the easily available prior information. 

\begin{table}[t]
\caption{Performance comparison of ConvLUT with different expert LUT numbers. 
 }\label{table:LUT}
\small
  \begin{center}
  \begin{tabular}{c|c|c|c}
    \toprule
     \multirow{2}{*}{\makecell[c]{LUT Number}} & \multirow{2}{*}{Storage} & \multirow{2}{*}{Runtime}& 1Mbps\\
     ~ & ~ & ~ & PSNR / SSIM\\
    \hline 
     1 & 2.32 MB& 8.64 ms& 24.21 / 0.6231\\
     3 & 4.84 MB& 9.31 ms& 24.60 / 0.6694\\
     6 & 8.65 MB& 10.23 ms& 24.84 / \textbf{0.6816}\\
     12& 16.27 MB& 12.05 ms& \textbf{24.89 } / 0.6813\\
    \bottomrule
  \end{tabular}
  \end{center}
\end{table}

\noindent\textbf{Configurations of Expert LUT.}
The Expert LUT of ConvLUT can be constructed from different SR networks. The last 3 rows of Table~\ref{table:SOTA} show how Expert LUT transferred from different SR networks affect the performance of ConvLUT, where better performed SISR network can help the corresponding Expert LUT to get better results in severe degradations. We also analyze how the number of Expert LUTs affects the performance of ConvLUT. We evenly selected $N$ QP values from the range of $0$ to $50$, where the corresponding videos are used to construct $N$ expert LUTs. As shown in Table~\ref{table:LUT}, increasing the number of expert LUTs constantly improves the SR performance, 
but more LUTs result in more storage and computational overhead. 
When the number of LUTs is larger than $6$, our model only has a minor improvement, and hence achieves the best trade-off between SR performance and efficiency. 

\begin{table}[t]
\caption{Runtime of LUT interpolation inference. After using our accelerated tetrahedral interpolation, the parallel acceleration can be applied without if-else control flow instructions. The inference speeds of both our model and SR-LUT get significantly improved.}\label{table:acc}
\small
  \renewcommand\tabcolsep{1pt} 
  \begin{center}
  
  \begin{tabular}{c|c|c|c}
    \toprule
    Model& Our Acceleration& If-else Control Flow  & Runtime\\
    \hline
    \multirow{2}{*}{SR-LUT~\cite{jo2021practical}} & \xmark & \cmark & 381.89 ms\\
     & \cmark & \xmark & \textbf{11.60 ms}\\
    \hline
    \multirow{2}{*}{ConvLUT} & \xmark & \cmark & 516.21 ms\\
    & \cmark & \xmark & \textbf{10.23 ms}\\
    \bottomrule
  \end{tabular}
\end{center}
\end{table}

\noindent\textbf{Effectiveness of Interpolation Acceleration.}
As shown in Table~\ref{table:acc}, with the proposed the efficient interpolation method, both SR-LUT~\cite{jo2021practical} and the proposed ConvLUT  achieve more than 35 times inference acceleration on GPU devices. Since ConvLUT only needs to query one fused LUT  while SR-LUT needs to repeatedly query one LUT for 4 times, our methods outperform SR-LUT in terms of runtime after acceleration. 

\noindent\textbf{Analysis of LUT Sampling.}
For LUT-based SR methods~\cite{jo2021practical,ma2022learning}, the original LUT is commonly sampled with a quantization value to compress the size of LUTs. For our method, we also uniformly sample the LUT. In table ~\ref{table:sampling}, we present the comparisons of our ConvLUT models with different quantization values. The uncompressed LUT bases ($2^0$) produces the best results but have unbearable storage (384GB). When the sampling size decreases from $2^2$ to $2^4$, the size of LUT significantly decreases from 1632MB to 7.644 MB while getting acceptable performance drop. Therefore, we choose the quantization value 16 as our default setting. If the LUT size matters, sampling sizes $2^5$ and $2^6$ could also be considered. For practical implementation, the sampling size should be considered as the tradeoff between the storage cost and the performance.

\begin{table}[t]
    \centering
    \caption{Comparison of our ConvLUT-RCAN with different sampling interval sizes. We set the sampling interval size as 16 for our model to reduce the LUT size, minimizing the drop of the original performance.}
    \begin{tabular}{l|c|c}
      \toprule
       \multirow{2}{*}{\makecell[c]{Sampling}} & LUT & 1Mbps\\
       ~ & Storage & PSNR / SSIM\\
      \hline 
       $2^0$ (Full LUT) & 384 GB & 24.91 / 0.6875\\
       $2^2$ & 1632 MB& 24.90 / 0.6871\\
       $2^3$ & 108 MB& 24.87 / 0.6843\\
       $2^4$ (Our) & 7.644 MB& 24.84 / 0.6816\\
       $2^5$ & 612 KB& 24.61 / 0.6740\\
       $2^6$ & 59.35 KB& 24.34 / 0.6538\\
       $2^8$ & 2.304 KB& 23.15 / 0.6459\\
      \bottomrule
    \end{tabular}
    \label{table:sampling}
  \vspace{-0.4cm}
  \end{table}

\section{Conclusion}
Online video streaming presents unique challenges for super-resolution due to dynamically changing degradations and strict latency requirements. 
This paper addresses this problem with a new benchmark dataset, LDV-WebRTC, produced with real-world online streaming system. A novel hybrid network that combines convolution and Look-Up Table (LUT) is proposed to achieve a better performance-latency trade-off. 
Our proposed mixture-of-expert-LUT module builds a set of LUTs specialized in different degradations and adaptively combines them to handle changing degradations. Experiment results show that our method achieves 720P video SR at around 100 FPS, outperforming existing LUT-based methods and offering competitive performance compared to efficient CNN-based methods.



\bibliographystyle{IEEEtran}
\bibliography{egbib}

\begin{thebibliography}{10}
\providecommand{\url}[1]{#1}
\csname url@samestyle\endcsname
\providecommand{\newblock}{\relax}
\providecommand{\bibinfo}[2]{#2}
\providecommand{\BIBentrySTDinterwordspacing}{\spaceskip=0pt\relax}
\providecommand{\BIBentryALTinterwordstretchfactor}{4}
\providecommand{\BIBentryALTinterwordspacing}{\spaceskip=\fontdimen2\font plus
\BIBentryALTinterwordstretchfactor\fontdimen3\font minus
  \fontdimen4\font\relax}
\providecommand{\BIBforeignlanguage}[2]{{%
\expandafter\ifx\csname l@#1\endcsname\relax
\typeout{** WARNING: IEEEtran.bst: No hyphenation pattern has been}%
\typeout{** loaded for the language `#1'. Using the pattern for}%
\typeout{** the default language instead.}%
\else
\language=\csname l@#1\endcsname
\fi
#2}}
\providecommand{\BIBdecl}{\relax}
\BIBdecl

\bibitem{lu2018you}
Z.~Lu, H.~Xia, S.~Heo, and D.~Wigdor, ``You watch, you give, and you engage: a
  study of live streaming practices in china,'' in \emph{Proceedings of the
  2018 CHI conference on human factors in computing systems}, 2018, pp. 1--13.

\bibitem{isobe2020revisiting}
T.~Isobe, F.~Zhu, X.~Jia, and S.~Wang, ``Revisiting temporal modeling for video
  super-resolution,'' \emph{BMVC}, 2020.

\bibitem{dong2015image}
C.~Dong, C.~C. Loy, K.~He, and X.~Tang, ``Image super-resolution using deep
  convolutional networks,'' \emph{IEEE transactions on pattern analysis and
  machine intelligence}, vol.~38, no.~2, pp. 295--307, 2015.

\bibitem{ledig2017photo}
C.~Ledig, L.~Theis, F.~Husz{\'a}r, J.~Caballero, A.~Cunningham, A.~Acosta,
  A.~Aitken, A.~Tejani, J.~Totz, Z.~Wang \emph{et~al.}, ``Photo-realistic
  single image super-resolution using a generative adversarial network,'' in
  \emph{Proceedings of the IEEE conference on computer vision and pattern
  recognition}, 2017, pp. 4681--4690.

\bibitem{ahn2018fast}
N.~Ahn, B.~Kang, and K.-A. Sohn, ``Fast, accurate, and lightweight
  super-resolution with cascading residual network,'' in \emph{Proceedings of
  the European conference on computer vision (ECCV)}, 2018, pp. 252--268.

\bibitem{zhang2018image}
Y.~Zhang, K.~Li, K.~Li, L.~Wang, B.~Zhong, and Y.~Fu, ``Image super-resolution
  using very deep residual channel attention networks,'' in \emph{Proceedings
  of the European conference on computer vision (ECCV)}, 2018, pp. 286--301.

\bibitem{caballero2017real}
J.~Caballero, C.~Ledig, A.~Aitken, A.~Acosta, J.~Totz, Z.~Wang, and W.~Shi,
  ``Real-time video super-resolution with spatio-temporal networks and motion
  compensation,'' in \emph{Proceedings of the IEEE conference on computer
  vision and pattern recognition}, 2017, pp. 4778--4787.

\bibitem{liu2022learning}
C.~Liu, H.~Yang, J.~Fu, and X.~Qian, ``Learning trajectory-aware transformer
  for video super-resolution,'' in \emph{Proceedings of the IEEE/CVF Conference
  on Computer Vision and Pattern Recognition}, 2022, pp. 5687--5696.

\bibitem{chan2022basicvsr++}
K.~C. Chan, S.~Zhou, X.~Xu, and C.~C. Loy, ``{BasicVSR++}: Improving video
  super-resolution with enhanced propagation and alignment,'' in
  \emph{Proceedings of the IEEE/CVF Conference on Computer Vision and Pattern
  Recognition}, 2022, pp. 5972--5981.

\bibitem{yang2020learning}
F.~Yang, H.~Yang, J.~Fu, H.~Lu, and B.~Guo, ``Learning texture transformer
  network for image super-resolution,'' in \emph{Proceedings of the IEEE/CVF
  conference on computer vision and pattern recognition}, 2020, pp. 5791--5800.

\bibitem{qiu2022learning}
Z.~Qiu, H.~Yang, J.~Fu, and D.~Fu, ``Learning spatiotemporal
  frequency-transformer for compressed video super-resolution,'' in
  \emph{Proceedings of the European conference on computer vision (ECCV)},
  2022.

\bibitem{johnston2012webrtc}
A.~B. Johnston and D.~C. Burnett, \emph{WebRTC: APIs and RTCWEB protocols of
  the HTML5 real-time web}.\hskip 1em plus 0.5em minus 0.4em\relax Digital
  Codex LLC, 2012.

\bibitem{yang2022ntire}
R.~Yang, R.~Timofte, M.~Zheng, Q.~Xing, M.~Qiao, M.~Xu, L.~Jiang, H.~Liu,
  Y.~Chen, Y.~Ben \emph{et~al.}, ``{NTIRE} 2022 challenge on super-resolution
  and quality enhancement of compressed video: Dataset, methods and results,''
  in \emph{Proceedings of the IEEE/CVF Conference on Computer Vision and
  Pattern Recognition}, 2022, pp. 1221--1238.

\bibitem{zeng2020learning}
H.~Zeng, J.~Cai, L.~Li, Z.~Cao, and L.~Zhang, ``Learning image-adaptive 3d
  lookup tables for high performance photo enhancement in real-time,''
  \emph{IEEE Transactions on Pattern Analysis and Machine Intelligence}, 2020.

\bibitem{wang2021real}
T.~Wang, Y.~Li, J.~Peng, Y.~Ma, X.~Wang, F.~Song, and Y.~Yan, ``Real-time image
  enhancer via learnable spatial-aware 3d lookup tables,'' in \emph{Proceedings
  of the IEEE/CVF International Conference on Computer Vision}, 2021, pp.
  2471--2480.

\bibitem{liu20224d}
C.~Liu, H.~Yang, J.~Fu, and X.~Qian, ``{4D LUT}: Learnable context-aware 4d
  lookup table for image enhancement,'' \emph{arXiv preprint arXiv:2209.01749},
  2022.

\bibitem{jo2021practical}
Y.~Jo and S.~J. Kim, ``Practical single-image super-resolution using look-up
  table,'' in \emph{Proceedings of the IEEE/CVF Conference on Computer Vision
  and Pattern Recognition}, 2021, pp. 691--700.

\bibitem{ma2022learning}
C.~Ma, J.~Zhang, J.~Zhou, and J.~Lu, ``Learning series-parallel lookup tables
  for efficient image super-resolution,'' in \emph{European Conference on
  Computer Vision}.\hskip 1em plus 0.5em minus 0.4em\relax Springer, 2022, pp.
  305--321.

\bibitem{li2022mulut}
J.~Li, C.~Chen, Z.~Cheng, and Z.~Xiong, ``{MuLUT}: Cooperating multiple look-up
  tables for efficient image super-resolution,'' in \emph{European Conference
  on Computer Vision}.\hskip 1em plus 0.5em minus 0.4em\relax Springer, 2022,
  pp. 238--256.

\bibitem{zhang2022clut}
F.~Zhang, H.~Zeng, T.~Zhang, and L.~Zhang, ``{CLUT-Net}: Learning adaptively
  compressed representations of {3DLUTs} for lightweight image enhancement,''
  in \emph{Proceedings of the 30th ACM International Conference on Multimedia},
  2022, pp. 6493--6501.

\bibitem{nvidia2018cuda}
C.~Nvidia, ``Cuda toolkit documentation,'' 2018.

\bibitem{wang2016performance}
Z.~Wang, B.~He, W.~Zhang, and S.~Jiang, ``A performance analysis framework for
  optimizing {OpenCL} applications on {FPGAs},'' in \emph{2016 IEEE
  International Symposium on High Performance Computer Architecture
  (HPCA)}.\hskip 1em plus 0.5em minus 0.4em\relax IEEE, 2016, pp. 114--125.

\bibitem{khorasani2015efficient}
F.~Khorasani, R.~Gupta, and L.~N. Bhuyan, ``Efficient warp execution in
  presence of divergence with collaborative context collection,'' in
  \emph{Proceedings of the 48th International Symposium on Microarchitecture},
  2015, pp. 204--215.

\bibitem{wang2004image}
Z.~Wang, A.~C. Bovik, H.~R. Sheikh, and E.~P. Simoncelli, ``Image quality
  assessment: from error visibility to structural similarity,'' \emph{IEEE
  transactions on image processing}, vol.~13, no.~4, pp. 600--612, 2004.

\bibitem{sullivan2012overview}
G.~J. Sullivan, J.-R. Ohm, W.-J. Han, and T.~Wiegand, ``Overview of the high
  efficiency video coding ({HEVC}) standard,'' \emph{IEEE Transactions on
  circuits and systems for video technology}, vol.~22, no.~12, pp. 1649--1668,
  2012.

\bibitem{yeo2020nemo}
H.~Yeo, C.~J. Chong, Y.~Jung, J.~Ye, and D.~Han, ``Nemo: enabling
  neural-enhanced video streaming on commodity mobile devices,'' in
  \emph{Proceedings of the 26th Annual International Conference on Mobile
  Computing and Networking}, 2020, pp. 1--14.

\bibitem{chen2020bitstream}
P.~Chen, W.~Yang, L.~Sun, and S.~Wang, ``When bitstream prior meets deep prior:
  Compressed video super-resolution with learning from decoding,'' in
  \emph{Proceedings of the 28th ACM International Conference on Multimedia},
  2020, pp. 1000--1008.

\bibitem{chan2021understanding}
K.~C. Chan, X.~Wang, K.~Yu, C.~Dong, and C.~C. Loy, ``Understanding deformable
  alignment in video super-resolution,'' in \emph{Proceedings of the AAAI
  conference on artificial intelligence}, vol.~35, no.~2, 2021, pp. 973--981.

\bibitem{chan2021basicvsr}
------, ``Basicvsr: The search for essential components in video
  super-resolution and beyond,'' in \emph{Proceedings of the IEEE/CVF
  Conference on Computer Vision and Pattern Recognition}, 2021, pp. 4947--4956.

\bibitem{yang2022adaint}
C.~Yang, M.~Jin, X.~Jia, Y.~Xu, and Y.~Chen, ``{AdaInt}: Learning adaptive
  intervals for 3d lookup tables on real-time image enhancement,'' in
  \emph{Proceedings of the IEEE/CVF Conference on Computer Vision and Pattern
  Recognition}, 2022, pp. 17\,522--17\,531.

\bibitem{xue2019video}
T.~Xue, B.~Chen, J.~Wu, D.~Wei, and W.~T. Freeman, ``Video enhancement with
  task-oriented flow,'' \emph{International Journal of Computer Vision}, vol.
  127, no.~8, pp. 1106--1125, 2019.

\bibitem{yang2021real}
X.~Yang, W.~Xiang, H.~Zeng, and L.~Zhang, ``Real-world video super-resolution:
  A benchmark dataset and a decomposition based learning scheme,'' in
  \emph{Proceedings of the IEEE/CVF International Conference on Computer
  Vision}, 2021, pp. 4781--4790.

\bibitem{wiegand2003overview}
T.~Wiegand, G.~J. Sullivan, G.~Bjontegaard, and A.~Luthra, ``Overview of the
  {H. 264/AVC} video coding standard,'' \emph{IEEE Transactions on circuits and
  systems for video technology}, vol.~13, no.~7, pp. 560--576, 2003.

\bibitem{jacobs1991adaptive}
R.~A. Jacobs, M.~I. Jordan, S.~J. Nowlan, and G.~E. Hinton, ``Adaptive mixtures
  of local experts,'' \emph{Neural computation}, vol.~3, no.~1, pp. 79--87,
  1991.

\bibitem{shazeer2017outrageously}
N.~Shazeer, A.~Mirhoseini, K.~Maziarz, A.~Davis, Q.~Le, G.~Hinton, and J.~Dean,
  ``Outrageously large neural networks: The sparsely-gated mixture-of-experts
  layer,'' \emph{arXiv preprint arXiv:1701.06538}, 2017.

\bibitem{emad2022moesr}
M.~Emad, M.~Peemen, and H.~Corporaal, ``{MoESR}: Blind super-resolution using
  kernel-aware mixture of experts,'' in \emph{Proceedings of the IEEE/CVF
  Winter Conference on Applications of Computer Vision}, 2022, pp. 3408--3417.

\bibitem{ulyanov2016instance}
D.~Ulyanov, A.~Vedaldi, and V.~Lempitsky, ``Instance normalization: The missing
  ingredient for fast stylization,'' \emph{arXiv preprint arXiv:1607.08022},
  2016.

\bibitem{maas2013rectifier}
A.~L. Maas, A.~Y. Hannun, A.~Y. Ng \emph{et~al.}, ``Rectifier nonlinearities
  improve neural network acoustic models,'' in \emph{ICML}, vol.~30,
  no.~1.\hskip 1em plus 0.5em minus 0.4em\relax Atlanta, Georgia, USA, 2013,
  p.~3.

\bibitem{kasson1995performing}
J.~M. Kasson, S.~I. Nin, W.~Plouffe, and J.~L. Hafner, ``Performing color space
  conversions with three-dimensional linear interpolation,'' \emph{Journal of
  Electronic Imaging}, vol.~4, no.~3, pp. 226--250, 1995.

\bibitem{zhao2020efficient}
H.~Zhao, X.~Kong, J.~He, Y.~Qiao, and C.~Dong, ``Efficient image
  super-resolution using pixel attention,'' in \emph{European Conference on
  Computer Vision}.\hskip 1em plus 0.5em minus 0.4em\relax Springer, 2020, pp.
  56--72.

\bibitem{cao2021video}
J.~Cao, Y.~Li, K.~Zhang, and L.~Van~Gool, ``Video super-resolution
  transformer,'' \emph{arXiv preprint arXiv:2106.06847}, 2021.

\bibitem{wang2019edvr}
X.~Wang, K.~C. Chan, K.~Yu, C.~Dong, and C.~Change~Loy, ``Edvr: Video
  restoration with enhanced deformable convolutional networks,'' in
  \emph{Proceedings of the IEEE/CVF Conference on Computer Vision and Pattern
  Recognition Workshops}, 2019, pp. 0--0.

\end{thebibliography}

\end{document}